\documentclass[a4paper,preprint,onecolumn,english,10pt]{elsarticle}

\usepackage[utf8x]{inputenc}
\usepackage{amsmath}
\usepackage{amssymb}
\usepackage[english]{babel}
\usepackage{graphicx}
\usepackage{epsfig}
\usepackage{verbatim}
\usepackage{latexsym}
\usepackage{color}
\usepackage{float}
\usepackage{caption}
\usepackage{subcaption}

\begin{document}

\begin{abstract}
We discuss the origin of multiscaling in financial time-series and investigate how to best quantify it.
Our methodology consists in separating the different sources of measured multifractality by analysing the multi/uni-scaling behaviour of synthetic time-series with known properties. We use  the results from the synthetic time-series to interpret the measure of multifractality of real log-returns time-series. The main finding is that the aggregation horizon of the returns can introduce a strong bias effect on the measure of multifractality. This effect can become especially important when returns distributions have power law tails with exponents in the range $[2,5]$. We discuss the right aggregation horizon to mitigate this bias.
\end{abstract}


\title{Measuring multiscaling in financial time-series}
\author[kings]{R. J. Buonocore}\ead{riccardo\_ junior.buonocore@kcl.ac.uk}
\author[ucl]{T. Aste}
\author[kings,ucl]{T. Di Matteo}
\address[kings]{Department of Mathematics, King's College London, The Strand, London WC2R 2LS, UK}
\address[ucl]{Department of Computer Science, University College London, Gower Street, London, WC1E 6BT, UK}
\date{}
\maketitle

\section{Introduction}
The multifractal behaviour of the financial time-series has become one the acknowledged stylized facts in the literature (see: \cite{mantegna_stanley_book,dacorogna_book,mantegna_stanley,scaling_review_tiziana,ramacont_review}). Many works have been dedicated to its empirical characterization \cite{ghashghaie,calvet1,liu1,bartolozzi,kristoufek,jiang}, reporting strong evidence of its presence in financial markets, and models \cite{mandelbrot1,calvet2,lux,lux_book,liu2,bacry1,bacry_other_models,bacry_skewed}.

Understanding which is the origin of the measured multifractality in financial markets is still an open research challenge.
This question has been raised first in \cite{kantelhardt} where the authors pointed out that the power law tails and the autocorrelation of the analysed time-series must be the two sources of the measured multifractality. 
In the first case, the multifractal behaviour is a consequence of the broadness of the unconditional distribution of the returns; while in the second case, the multifractal behaviour is associated with the causal structure of the time-series. 
After \cite{kantelhardt}, many papers have investigated the relative contribution of these two sources to the measured multifractality, however no agreement exists. For example in \cite{zhou} the author points out that the autocorrelation structure has a minor impact on the measured multifractality while the power law tails are the major source of it. In \cite{barunik} they also report that the power law tails give the major contribution, but they also point out that the presence of unknown autocorrelations might introduce a negative bias effect in the quantification of multifractality. Conversely, in \cite{green} the authors find that the autocorrelation gives the major contribution while for a specific time-series  the ``extreme events are actually inimical to the multifractal scaling''. This lack of agreement motivated our work, leading us to investigate what the source of the measured multifractality is and how it can be detected.

In this paper we quantify the two contributions by using synthetic times series where the two contributions can be separated.  Specifically we analyze Brownian Motion with innovations drawn from a t-Student distribution,  Multifractal Random Walk and normalized version of the Multifractal Random Walk. The measured multifractality on these synthetic series are compared with measures on both real financial log-returns and on a normalized version of the real log-returns where the heavy tails are removed. The results show the aggregation horizon has a strong effect on the quantification of multifractality. We verify however that there are regions of the aggregation horizon that can be used in practice to extract reliable multifractality estimators.

The rest of the paper is organized as follows: in Sec. \ref{background} we perform a brief literature review introducing the tools we used for our analysis and discussing the results from  previous works. In Sec. \ref{section_methods} we review the theoretical models we used and we define the multifractality estimators that shall be used throughout the paper. Secs. \ref{sec_source} and \ref{section_real} are dedicated respectively to the analysis of artificial and real data. In Sec. \ref{section_discussion} we discuss the results while in Sec. \ref{section_summary} we summarize the results and conclude.

\section{Background}\label{background}

\subsection{Multifractality}\label{subsec_multifractality}
Among the methods which are used for the empirical measurement of the scaling exponents, in this work we will use only the \textit{Generalized Hurst Exponent method} (GHE), see \cite{scaling_review_tiziana,kantelhardt,tiziana_dacorogna2}\footnote{The code can be found at http://www.mathworks.com/matlabcentral/fileexchange/30076-generalized-hurst-exponent.} which relies on the measurement of the direct scaling of the $q$th-order moments of the distribution of the increments and it has been shown to be one of the most reliable estimators \cite{barunik_tails}. Let us call $X(t)$ a process with stationary increments. The GHE method considers the following function of the increments
\begin{equation}\label{moments_scaling}
E[|X(t+\tau)-X(t)|^q]=K(q)\tau^{qH(q)},
\end{equation}
where $\tau$ is the time horizon over which the increments are computed and $H(q)$ is the Generalized Hurst Exponent. The function $\zeta(q)=qH(q)$ is concave and $K(q)$ depends also on $q$. In particular, GHE considers the logarithm of Eq. (\ref{moments_scaling})
\begin{equation}\label{log_moments_scaling2}
\ln\left(E[|X(t+\tau)-X(t)|^q]\right)=\zeta(q)\ln(\tau)+\ln\left(K(q)\right),
\end{equation}
and, if linearity with respect to $\ln(\tau)$ holds, it computes the slopes of the straight lines at different $q$. The slopes are computed in the following way: for every $q$, several linear fits are computed taking $\tau\in[\tau_{min},\tau_{max}]$, with usually $\tau_{min}=1$ and several values of $\tau_{max}$ typically between $[5,19]$; the output estimator for $\zeta(q)$ is the average of these values for a given $q$. This method gives also the errors which are the standard deviations of these values. However, in this paper we do not perform any average over  different  values of $\tau_{min}$, $\tau_{max}$  and we instead consider just one linear fit for a given range $\tau\in[\tau_{min},\tau_{max}]$. In particular we focus our attention on two ranges, namely $\tau\in[1,19]$, following the prescription of other works (\cite{tiziana_dacorogna2,tiziana_dacorogna,morales1}), and $\tau\in[30,250]$. The reason for this simplification is that, given a range of $\tau$, we did not want to weight more the small values with respect to the big values. This point will be further stressed later in the paper.

\subsection{Source of multiscaling in financial data: state of the art}
As already mentioned in the Introduction, there is a debate in literature concerning what property of the financial time-series contributes mostly to their observed multiscaling behaviour. Let us here discuss some findings present in the literature. In \cite{zhou} the author studied the Dow Jones Industrial Average taken on a daily basis and processed the data in four different ways in order to uncover the source of the multiscaling behaviour. The methods used were (\cite{zhou}):
\begin{enumerate}
\item shuffling the data in order to check the impact of the shape of the unconditional distribution;
\item building up surrogate data with the same unconditional distribution and linear correlation of the empirical one but with any non linear correlation removed;
\item cutting the tails by substituting the more extreme events with resampled ones from the core of the distribution;
\item generating surrogate power law-tailed time-series, namely double Weibull and t-Student, preserving the temporal structure of the empirical time-series.
\end{enumerate}
The author found that, on one hand the temporal structure, both linear and non linear, has a minor impact. On the other hand, the fatter the tails are, the stronger the multiscaling. And this result was confirmed both by cutting the extreme events and changing the unconditional distribution.

In \cite{green} the authors studied again the Dow Jones Industrial Average taken on a daily basis plus the Dow Jones Euro Stoxx 50 sampled at one minute. In this case three analysis were performed:
\begin{enumerate}
\item shuffling the whole dataset;
\item dividing the dataset into intervals and shuffling them in order to keep short memory contributions then repeating the analysis changing the length of the intervals;
\item cutting the extreme events.
\end{enumerate}
The authors found that when shuffled, the dataset loses its multiscaling behaviour (\cite{green}). The shuffling of the intervals showed that the linearity of the scaling of the fluctuation functions worsen when the length of the interval is small and improves increasing it, thus according to the authors this should be regarded as a sign that the temporal correlations are the source of multiscaling. For what concerns the cut of the most extreme events they found that for the Dow Jones Industrial Average extreme events have no particular impact, while for the Dow Jones Euro Stoxx 50 they cause a distortion in the Singularity Spectrum.

Finally in \cite{barunik} an extensive analysis was conducted on several empirical time-series including stock market indexes, exchange rates and interest rates. In order to unveil the source of the empirical multiscaling, the shuffling method was used plus a comparison with synthetic data. The authors also found an increase of the measured multiscaling of the shuffled time-series which then led them to draw two conclusions: first that the major source of the multifractality comes from the power law tails of the distribution; second that the presence of time correlations decreases the multifractality. These conclusions are consistent with the analysis of the Markov Switching Multifractal Model (\cite{lux}). Further analyses have been conducted by means of fractional Brownian motions, random walks with steps drawn from a Levy distribution and ARFIMA processes, all confirming the results found on the empirical datasets (\cite{barunik}).

\section{Models and methods}\label{section_methods}
In this section we describe the analytical properties of the models we used for our analysis and the variable we chose to detect the multifractality.
\subsection{Brownian motion with t-Student innovations (tBM)}
We considered a uniscaling process with independent increments drawn from a t-Student distribution. Introducing the dummy variable $t$, the probability density of a t-Student distribution is given by (\cite{tstud})
\begin{equation}\label{t-Stud_classic}
p(t)=\frac{\Gamma(\frac{n+1}{2})}{\sqrt{n\pi}\Gamma(\frac{n}{2})}\left(1+\frac{t^2}{n}\right)^{-\left(\frac{n+1}{2}\right)},
\end{equation}
where $n$ is the number of degrees freedom which can be non-integer. According to Eq. (\ref{t-Stud_classic}) the variable $t$ has mean zero  if $n>1$ and infinite otherwise. The variance is instead equal to $\displaystyle \frac{n}{n-2}$ if $n>2$, infinite if $1<n<2$ and undefined otherwise. The spectrum of a tBM can be computed analytically in both cases, either if $n$ is bigger or smaller then two. For $n<2$ the t-Student distribution of Eq. (\ref{t-Stud_classic}) behaves as a stable distribution with skewness parameter equal to zero and stability parameter equal to $n$, so the scaling exponents are (see \cite{kantelhardt,nakao,chechkin})\footnote{In \cite{kantelhardt} is reported the shape of the scaling exponent for $q>n$ to be equal to one. However, as underlined in \cite{nakao} and \cite{chechkin}, this so called bifractal behaviour is a pure finite size sample effect.}
\begin{equation}\label{scaling_t_stud_2}
\zeta(q)=qH(q)=\frac{q}{n}\quad\mbox{if $q<n$}.
\end{equation}
For $n>2$ and finite aggregation horizon $\tau$ it can be shown that
\begin{equation}
E[|X(t+\tau)-X(t)|^q]=f(q)\tau^\frac{q}{2}.
\end{equation}
Thus
\begin{equation}\label{scaling_t_stud}
\zeta(q)=qH(q)=\frac{q}{2}\quad\mbox{if $q<n$}.
\end{equation}
It is expected then that for $n>2$ the scaling exponents are identical to the one of a BM up to $q=n$. For $n=2$, it can be proved rigorously that the scaling exponents behave like Eq. (\ref{scaling_t_stud}) (cfr. \cite{tStud2}).

According to these analytical observations a tBM is a unifractal process both for $n<2$ and $n\geq2$ and $\zeta(q)$ behaves as a straight line.

\subsection{Multifractal Random Walk (MRW)}\label{section_mrw}
Among the models proposed in the literature, in the present paper we chose as a benchmark multifractal model the so-called Multifractal Random Walk introduced in \cite{bacry1}. Its main appealing property is that it has exactly computable scaling exponents. We report that this model has been further developed and alternative multifractal random walks models with different scaling exponents has been proposed (see \cite{bacry_other_models,bacry_skewed}), however for our purposes the statistical properties of this original model are sufficient. In the discrete version, the process $X(t)$ described by the model is defined as (\cite{bacry1})
\begin{equation}
X(t)=\sum_{k=1}^{\frac{t}{\Delta t}}\epsilon_{\Delta t}(k)e^{\omega_{\Delta t}(k)},
\end{equation}
so the increments can be written as
\begin{equation}
r_\tau(t)=X(t+\tau)-X(t)=\sum_{k=\frac{t}{\Delta t}+1}^{\frac{t+\tau}{\Delta t}}\epsilon_{\Delta t}(k)e^{\omega_{\Delta t}(k)},
\end{equation}
with $\epsilon_{\Delta t}\sim N(0,\sigma^2\Delta t)$, $\omega_{\Delta t}\sim N(-\lambda^2\ln(L/\Delta t),\lambda^2\ln(L/\Delta t))$, where $\lambda$ is called intermittency parameter, $L$ is the autocorrelation length, $\sigma$ is the variance of the overall process and $\Delta t$ is the discretization step (\cite{bacry1}). The peculiarity of this model is that, while the $\epsilon_{\Delta t}(k)$ are independent, the $\omega_{\Delta t}(k)$ are not, having autocovariance (\cite{bacry1}):
\begin{equation}
Cov(\omega_{\Delta t}(k_1),\omega_{\Delta t}(k_2))=\lambda^2\ln\rho_{\Delta t}(k_1-k_2),
\end{equation}
with
\begin{equation}
\rho_{\Delta t}(k_1-k_2)=
\left\{
\begin{array}{cc}
\displaystyle
\frac{L}{(|k_1-k_2|+1)\Delta t} & |k_1-k_2|<L/\Delta t,\\
1 & \mbox{otherwise}.
\end{array}
\right.
\end{equation}
The scaling exponents of this model in the continuous limit are (\cite{bacry1}):
\begin{equation}\label{spectrum_MRW}
\zeta(q)=qH(q)=-\frac{\lambda^2}{2}q^2+(\lambda^2+\frac{1}{2})q.
\end{equation}
The importance of this model relies in the fact that, by means of just three parameters ($\lambda,L,\sigma$), it exhibits both power law tails and volatility clustering, keeping its plain innovations uncorrelated. In particular the intermittency parameter $\lambda$ determines both the power law tails, which decay with an exponent proportional to $\lambda^2$ (\cite{bacry2}), and the decay of the autocorrelation functions of the powers of the absolute returns, whose decaying exponents are again proportional to $\lambda^2$ (\cite{bacry1}).

\subsection{Multifractality estimator}\label{sec_estimator_choice}
In order to understand the behaviour of the scaling exponents $\zeta(q)$, we used the Generalized Hurst Exponent, $H(q)$ (see Eqs. (\ref{moments_scaling}) and (\ref{log_moments_scaling2})). Let us note that due to the presence of the power law tails in the empirical datasets (\cite{ramacont_review,chakraborti_review}), the value of $q$ should be less then the tail exponent of the analysed time-series, since the moments are not finite for large $q$. Moreover, the existence of a moment does not guarantee its measurement on finite samples to be reliable when its variance is not finite. Following these observations, along with the fact that the decay exponents of the empirical power law tails typically range between two and five (\cite{ramacont_review}), in our analyses we limited ourselves to $q\leq1$. In particular we took a range of $q$ between $0.1$ and $1$ every $0.1$ units, having $10$ points in total\footnote{We checked that increasing the number of points over the interval does not change the results.}.
\\
To assess the presence of a statistically meaningful curvature in the scaling exponents, thus multiscaling, we performed a parabolic fit over the range $q\leq1$ and then we took the coefficient of the second degree term as a multiscaling estimator, \textit{i.e.}
\begin{equation}\label{new_estimator}
\zeta(q)=qH(q)\simeq Bq^2+Aq+const,
\end{equation}
where then $\hat B$ is the multifractality estimator\footnote{The notation of the hat means the estimator of the quantity under it.} we adopted in this paper. It must have negative (multiscaling behaviour) or zero (uniscaling behaviour) expectation value (due to concavity). The expected value of the parameter $const$ is zero and in our measurements of $\zeta(q)$ we always checked this condition for consistency. Note that in \cite{stanley_3} the authors fit the Singularity Spectrum, with a fourth degree polynomial which implies necessarily a fourth degree polynomial functional form for $\zeta(q)$. However, for our purposes, a second degree fit is enough and we verified that the inclusion of the terms up to the fourth degree does not modify our results.

\section{Analysis of artificial data}\label{sec_source}
We started our analysis simulating $10^4$ MRW processes, specified in Subsec. \ref{section_mrw}, made of $10^6$ steps $\delta t$ with parameters $\lambda^2=0.03$, $L=1000$, $\sigma=1$ and computing the mean and the standard deviation of $\hat B$, $\hat H(0.5)$ and $\hat H(1)$ over the realizations. We then repeated the measure over the shuffled version of the time-series. The convergence of the estimators has been always checked. The values of $\lambda^2$ and $L$ have been chosen according to empirical analyses conducted in other works (see for example \cite{morales2}), while the length has been chosen to reduce as much as possible the finite size sample errors keeping reasonable computational times. The results are reported in Tabs. \ref{plainVSshuffledMRW1} and \ref{plainVSshuffledMRW2} with respectively $\tau\in[1,19]$ and $\tau\in[30,250]$. The theoretical values are reported in boldface within brackets under the measured values.
\begin{table}[H]
\begin{center}
\begin{tabular}{|c|c|c|}
\hline
MRW				&								plain								&							shuffled							\\
\hline
$\hat B$		&$\begin{matrix}-0.0090\pm0.0006 \\ \mathbf{(-0.015)}\end{matrix}$	&$\begin{matrix}-0.0273\pm0.0006 \\ \mathbf{(0)}\end{matrix}$\\
\hline
$\hat H(0.5)$	&$\begin{matrix}0.514\pm0.001 \\ \mathbf{(0.5225)}\end{matrix}$	&$\begin{matrix}0.541\pm0.001 \\ \mathbf{(0.5)}\end{matrix}$\\
\hline
$\hat H(1)$		&$\begin{matrix}0.509\pm0.001 \\ \mathbf{(0.515)}\end{matrix}$		&$\begin{matrix}0.527\pm0.001 \\ \mathbf{(0.5)}\end{matrix}$\\
\hline
\end{tabular}
\caption{Comparison between $\hat B$, $\hat H(0.5)$ and $\hat H(1)$ for a plain and a shuffled MRW with $\tau\in[1,19]$.}
\label{plainVSshuffledMRW1}
\end{center}
\end{table}

\begin{table}[H]
\begin{center}
\begin{tabular}{|c|c|c|}
\hline
MRW				&								plain								&							shuffled							\\
\hline
$\hat B$		&$\begin{matrix}-0.014\pm0.002 \\ \mathbf{(-0.015)}\end{matrix}$	&$\begin{matrix}-0.002\pm0.002 \\ \mathbf{(0)}\end{matrix}$\\
\hline
$\hat H(0.5)$	&$\begin{matrix}0.521\pm0.005 \\ \mathbf{(0.5225)}\end{matrix}$	&$\begin{matrix}0.503\pm0.005 \\ \mathbf{(0.5)}\end{matrix}$\\
\hline
$\hat H(1)$		&$\begin{matrix}0.514\pm0.005 \\ \mathbf{(0.515)}\end{matrix}$		&$\begin{matrix}0.502\pm0.005 \\ \mathbf{(0.5)}\end{matrix}$\\
\hline
\end{tabular}
\caption{Comparison between $\hat B$, $\hat H(0.5)$ and $\hat H(1)$ for a plain and a shuffled MRW with $\tau\in[30,250]$.}
\label{plainVSshuffledMRW2}
\end{center}
\end{table}

It is evident from the Tables that in the region $\tau\in[1,19]$ also for MRW the non linearity of the scaling exponents increases after shuffling confirming the results of \cite{barunik}, while in the region $\tau\in[30,250]$ this effect disappears and the shuffled process seems statistically undistinguishable from a BM. According to its definition (see Sec. \ref{background}), a shuffled MRW is an uncorrelated, symmetric time-series with power law tails. In light of this, a model which might give us some further indication is a tBM. In the next subsection we focus on this model.

\subsection{The effect of the power law tails}\label{subsec_fat_tails}
\begin{figure}[H]
\begin{center}
\includegraphics[scale=0.953]{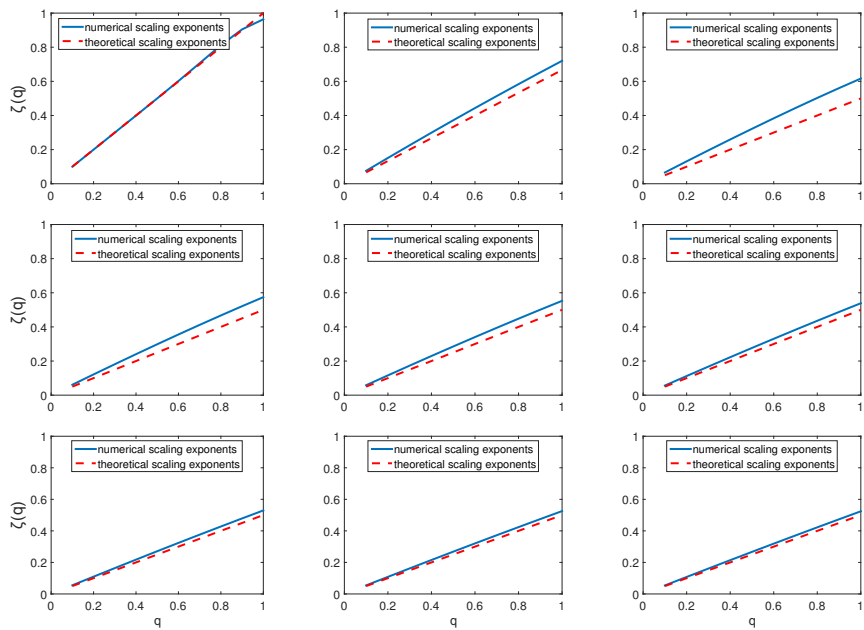}
\end{center}
\caption{Numerical values of $\zeta(q)$ (blue line) against its theoretical values (red dashed line) for a tBM with $n=[1,1.5,2,2.5,3,3.5,4,4.5,5]$ taken every $0.5$ units, in increasing order from left to right and top to bottom.}
\label{figure_t_stud}
\end{figure}

Let us here report the estimators $\hat B$, $\hat H(0.5)$ and $\hat H(1)$ in the presence of power law tails. In Fig. \ref{figure_t_stud} we report the results of the computation of the scaling exponents $\zeta(q)$ for $\tau\in[1,19]$ of single realizations of processes with t-Student innovations made of $10^6$ steps, for various values of $n$: $n\in[1,5]$ every $0.5$ units (cfr. Eq. \ref{t-Stud_classic}). In blue solid line the measured scaling exponents of the synthetic time-series are reported, whereas in dashed red line the theoretical expectation (see Eqs. (\ref{scaling_t_stud_2}) and (\ref{scaling_t_stud})). It is evident that as soon as $n$ moves away from $1$, a curvature of $\zeta(q)$ arises. But it is also evident that, as the tail index increases above $n=2$ the graphs become more linear with apparent linearity almost recovered above $n=5$. It is worth noting that the empirically measured tail indexes fall exactly in the range $[2,5]$, which is the most numerically biased. In order make a quantitative assessment, for each value of $n=3,4,5$, which roughly covers the range of empirically observed tails, we simulated $10^4$ tBM made of $10^6$ steps and we computed the mean and the standard deviation of $\hat B$, $\hat H(0.5)$ and $\hat H(1)$ for all values of $n$. Tabs. \ref{B-VS-tstud1} and \ref{B-VS-tstud2} report the numerical results for respectively $\tau\in[1,19]$ and $\tau\in[30,250]$ (theoretical values in boldface within brackets under measured values).
\begin{table}[H]
\begin{center}
\begin{tabular}{|c|c|c|c|}
\hline 
tBM				& 							$n=3$							&							$n=4$								&							$n=5$								\\ 
\hline
$\hat B$		&$\begin{matrix}-0.0364\pm0.0007 \\ \mathbf{(0)}\end{matrix}$&	$\begin{matrix}-0.0251\pm0.0005 \\ \mathbf{(0)}\end{matrix}$&$\begin{matrix}-0.0186\pm0.0005 \\ \mathbf{(0)}\end{matrix}$\\ 
\hline 
$\hat H(0.5)$	&$\begin{matrix}0.570\pm0.001 \\ \mathbf{(0.5)}\end{matrix}$&$\begin{matrix}0.544\pm0.001 \\ \mathbf{(0.5)}\end{matrix}$	&$\begin{matrix}0.531\pm0.001 \\ \mathbf{(0.5)}\end{matrix}$\\
\hline 
$\hat H(1)$		&$\begin{matrix}0.552\pm0.001 \\ \mathbf{(0.5)}\end{matrix}$&$\begin{matrix}0.531\pm0.001 \\ \mathbf{(0.5)}\end{matrix}$	&$\begin{matrix}0.522\pm0.001 \\ \mathbf{(0.5)}\end{matrix}$\\
\hline
\end{tabular} 
\caption{Mean and standard deviation of $\hat B$, $\hat H(0.5)$ and $\hat H(1)$ computed on t-Students time-series with $n=3,4,5$ and $\tau\in[1,19]$.}
\label{B-VS-tstud1}
\end{center}
\end{table}

\begin{table}[H]
\begin{center}
\begin{tabular}{|c|c|c|c|}
\hline 
tBM				& 							$n=3$							&							$n=4$								&							$n=5$								\\ 
\hline
$\hat B$		&$\begin{matrix}(-9\pm2)\cdot10^{-3} \\ \mathbf{(0)}\end{matrix}$&	$\begin{matrix}(-4\pm2)\cdot10^{-3} \\ \mathbf{(0)}\end{matrix}$&$\begin{matrix}(-2\pm2)\cdot10^{-3} \\ \mathbf{(0)}\end{matrix}$\\ 
\hline 
$\hat H(0.5)$	&$\begin{matrix}0.517\pm0.005 \\ \mathbf{(0.5)}\end{matrix}$&$\begin{matrix}0.506\pm0.005 \\ \mathbf{(0.5)}\end{matrix}$	&$\begin{matrix}0.503\pm0.005 \\ \mathbf{(0.5)}\end{matrix}$\\
\hline 
$\hat H(1)$		&$\begin{matrix}0.513\pm0.005 \\ \mathbf{(0.5)}\end{matrix}$&$\begin{matrix}0.504\pm0.005 \\ \mathbf{(0.5)}\end{matrix}$	&$\begin{matrix}0.502\pm0.005 \\ \mathbf{(0.5)}\end{matrix}$\\
\hline
\end{tabular} 
\caption{Mean and standard deviation of $\hat B$, $\hat H(0.5)$ and $\hat H(1)$ computed on t-Students time-series with $n=3,4,5$ and $\tau\in[30,250]$.}
\label{B-VS-tstud2}
\end{center}
\end{table}
Let us note that in the range $\tau\in[1,19]$ with a significance level of $1\%$, a multiscaling behaviour is found due to the presence of power law tails in all cases, while in the range $\tau\in[30,250]$ only the case $n=3$, keeps its concavity at $1\%$ significance level, but still very lowered with respect to the other region. Thus the measurements in the latter region seem to agree better with the theoretical uniscaling behaviour.

\subsection{Effect of autocorrelations}\label{subsec_autocorrelation}
\begin{figure}[!ht]
\begin{center}
\includegraphics[scale=0.4416]{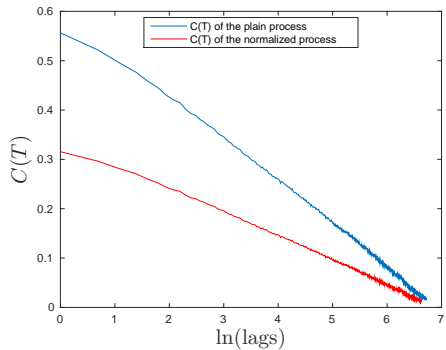}
\caption{Autocovariance function of the log absolute returns for a plain (blue) and normalized (red) path drawn from a MRW made of $10^6$ steps with $\lambda=0.3$, $L=1000$, $\sigma=1$.}
\label{norm_autocorr}
\end{center}
\end{figure}

\begin{figure}[H]
   \begin{center}
   \includegraphics[width=1\textwidth]{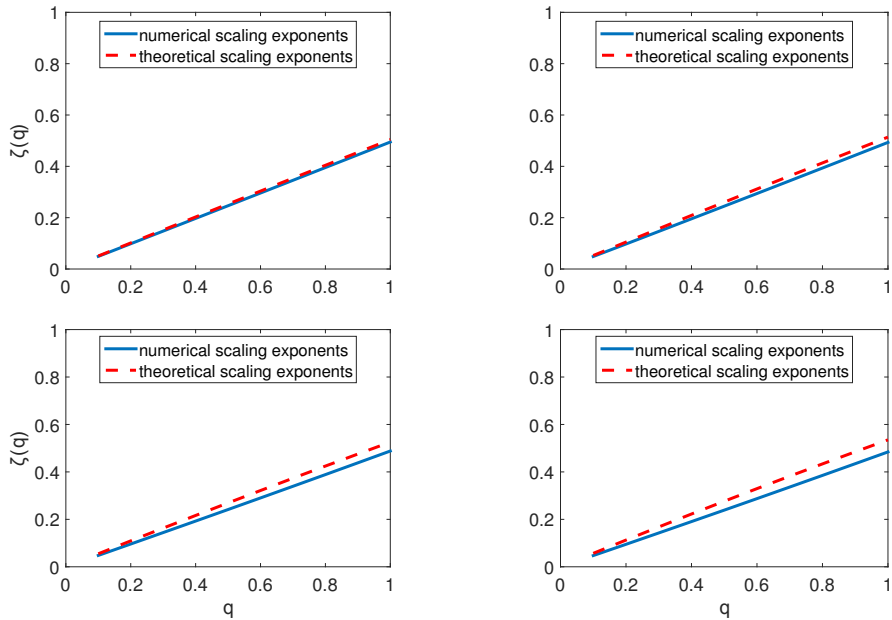}
   \end{center}
    \caption{Numerical values of $\zeta(q)$ (blue line) against its theoretical values (red dashed line) for a nMRW with $\lambda=[0.1,0.2,0.3,0.4]$, in increasing order from left to right and top to bottom.}
    \label{figure_nmrw}
\end{figure}

In order to isolate the contribution of the autocorrelation and eliminate the effect of the tails, we applied a normalization procedure to the MRW. The method consists in changing the unconditional distribution of a time-series into a desired one preserving its causal structure as proposed in \cite{zhou}. Before we proceed we need to stress a detail. If the empirical time series has power law tails while the surrogate is normally distributed, the autocovariance of the second one has the same functional form of the first one, but its strength is lowered. This can be simply ascribed to the fact that the extreme events give a big contribution in the computations of the averages, thus normalizing them reduces the strength of the correlations at each lag. This effect can be easily seen by plotting in semilog scale on the same figure the function proposed in \cite{bacry2} for the estimation of the parameters of the MRW for a MRW and its normalized version (nMRW). This is shown in Fig. \ref{norm_autocorr} in semilog scale where we observe that the autocovariance of the original time-series follows well the theoretical behaviour (\cite{bacry2})
\begin{equation}\label{C_T}
C(T)=Cov\left[\ln|r_\tau(t+T)|,\ln|r_\tau(t)|\right]=\lambda^2\ln\left(\frac{L}{T+1}\right),
\end{equation}
whereas the normalized one has a smaller effective value of $\lambda$. It is evident that the slope of the line relative to the normalized process is smaller than the slope of the line relative to the plain one (in absolute value).
The behaviours of the scaling exponents $\zeta(q)$ for $\tau\in[1,19]$ of single realizations of nMRWs made of $10^6$ steps for different degree of autocorrelation $\lambda$, specified in the captions, $L=1000$ and $\sigma=1$ are reported in Fig. \ref{figure_nmrw}. As noted before the effective value of $\lambda$ after the normalization is a bit lower then the one reported in the captions, so the theoretical line is plotted recomputing the value of $\lambda$ over the normalized processes.

We observe that, in all cases, the function $\zeta(q)$ changes its concavity. In order to make a quantitative assessment, for each value of $\lambda^2=0.03,0.04,0.05$, we simulated $10^4$ MRWs made of $10^6$ steps, we normalized them and we computed the mean and the standard deviation of $\hat B$ along with $\hat H(0.5)$ and $\hat H(1)$. Tabs. \ref{B-VS-nMRW1} and \ref{B-VS-nMRW2} report the numerical results for $\tau\in[1,19]$ and $\tau\in[30,250]$ together with the theoretical expected values in boldface under the measured one. The effective value of $\lambda$, called $\lambda_{eff}$ in the table, which affects $\hat B$, $\hat H(0.5)$ and $\hat H(1)$, was obtained from Eq. (\ref{C_T}) by fitting the autocovariance of each normalized time-series taking then the mean and the standard deviation.
\begin{table}[H]
\begin{center}
\begin{tabular}{|c|c|c|c|}
\hline 
				 &							$\lambda^2=0.03$								&									$\lambda^2=0.04$						&						$\lambda^2=0.05$ \\ 
\hline 
$\lambda^2_{eff}$& 					$0.0223\pm0.0005$										&							$0.0279\pm0.0007$								&							$0.0330\pm0.0008						$\\ 
\hline 
$\hat B$		 &$\begin{matrix}0.0075\pm0.0005 \\ \mathbf{(-0.0111\pm0.0003)}\end{matrix}$&$\begin{matrix}0.0085\pm0.0005 \\ \mathbf{(-0.0139\pm0.0003)}\end{matrix}$	&$\begin{matrix}0.0093\pm0.0006 \\ \mathbf{(-0.0165\pm0.0004)}\end{matrix}$\\
\hline 
$\hat H(0.5)$	 &$\begin{matrix}0.489\pm0.001 \\ \mathbf{(0.5167\pm0.0004)}\end{matrix}$	&$\begin{matrix}0.487\pm0.001 \\ \mathbf{(0.5209\pm0.0005)}\end{matrix}$	&$\begin{matrix}0.486\pm0.001 \\ \mathbf{(0.5247\pm0.0006)}\end{matrix}$\\ 
\hline 
$\hat H(1)$		 &$\begin{matrix}0.492\pm0.001 \\ \mathbf{(0.5111\pm0.0003)}\end{matrix}$	&$\begin{matrix}0.491\pm0.001 \\ \mathbf{(0.5139\pm0.0003)}\end{matrix}$	&$\begin{matrix}0.490\pm0.001 \\ \mathbf{(0.5165\pm0.0004)}\end{matrix}$\\ 
\hline 
\end{tabular}
\caption{Mean and standard deviation of $\hat B$, $\hat H(0.5)$ and $\hat H(1)$ computed on nMRWs with $L=1000$, $\sigma=1$ and $\tau\in[1,19]$.}
\label{B-VS-nMRW1}
\end{center}
\end{table}

\begin{table}[H]
\begin{center}
\begin{tabular}{|c|c|c|c|}
\hline 
				 &							$\lambda^2=0.03$								&									$\lambda^2=0.04$						&						$\lambda^2=0.05$ \\ 
\hline 
$\lambda^2_{eff}$&							$0.0223\pm0.0005$								&									$0.0279\pm0.0006$						&						$0.0330\pm0.0008$							\\ 
\hline 
$\hat B$		 &$\begin{matrix}-0.007\pm0.002 \\ \mathbf{(-0.0111\pm0.0003)}\end{matrix}$	&$\begin{matrix}-0.009\pm0.002 \\ \mathbf{(-0.0139\pm0.0003)}\end{matrix}$	&$\begin{matrix}-0.010\pm0.002 \\ \mathbf{(-0.0165\pm0.0004)}\end{matrix}$\\
\hline 
$\hat H(0.5)$	 &$\begin{matrix}0.511\pm0.005 \\ \mathbf{(0.5167\pm0.0004)}\end{matrix}$		&$\begin{matrix}0.513\pm0.005 \\ \mathbf{(0.5209\pm0.0005)}\end{matrix}$		&$\begin{matrix}0.515\pm0.005 \\ \mathbf{(0.5247\pm0.0006)}\end{matrix}$\\ 
\hline 
$\hat H(1)$		 &$\begin{matrix}0.507\pm0.005 \\ \mathbf{(0.5111\pm0.0003)}\end{matrix}$		&$\begin{matrix}0.508\pm0.005 \\ \mathbf{(0.5139\pm0.0003)}\end{matrix}$		&$\begin{matrix}0.509\pm0.005 \\ \mathbf{(0.5165\pm0.0004)}\end{matrix}$\\ 
\hline 
\end{tabular}
\caption{Mean and standard deviation of $\hat B$, $\hat H(0.5)$ and $\hat H(1)$ computed on nMRWs with $L=1000$, $\sigma=1$ and $\tau\in[30,250]$.}
\label{B-VS-nMRW2}
\end{center}
\end{table}

These results confirm the change of the concavity of the scaling exponents in the region $\tau\in[1,19]$. Indeed, we observe in Tab. \ref{B-VS-nMRW2} that, within the $1\%$ significance level, all $\hat B$ stay positive. Positive values of $\hat B$ imply the convexity of the function $\zeta(q)$, which, in the multifractal picture, is supposed to be concave.\\
The region $\tau\in[30,250]$ is instead much more well-behaved having in all three cases concave scaling exponents within the $1\%$ significance level, despite for $\lambda^2=0.05$ only (the most correlated) the measured $\hat B$ falls slightly outside the $1\%$ significance level from the expected value.

\section{Analysis of a real dataset}\label{section_real}
\subsection{Dataset}
\begin{figure}[H]
\begin{center}
\begin{minipage}[b]{0.49\textwidth}
\includegraphics[width=1\textwidth]{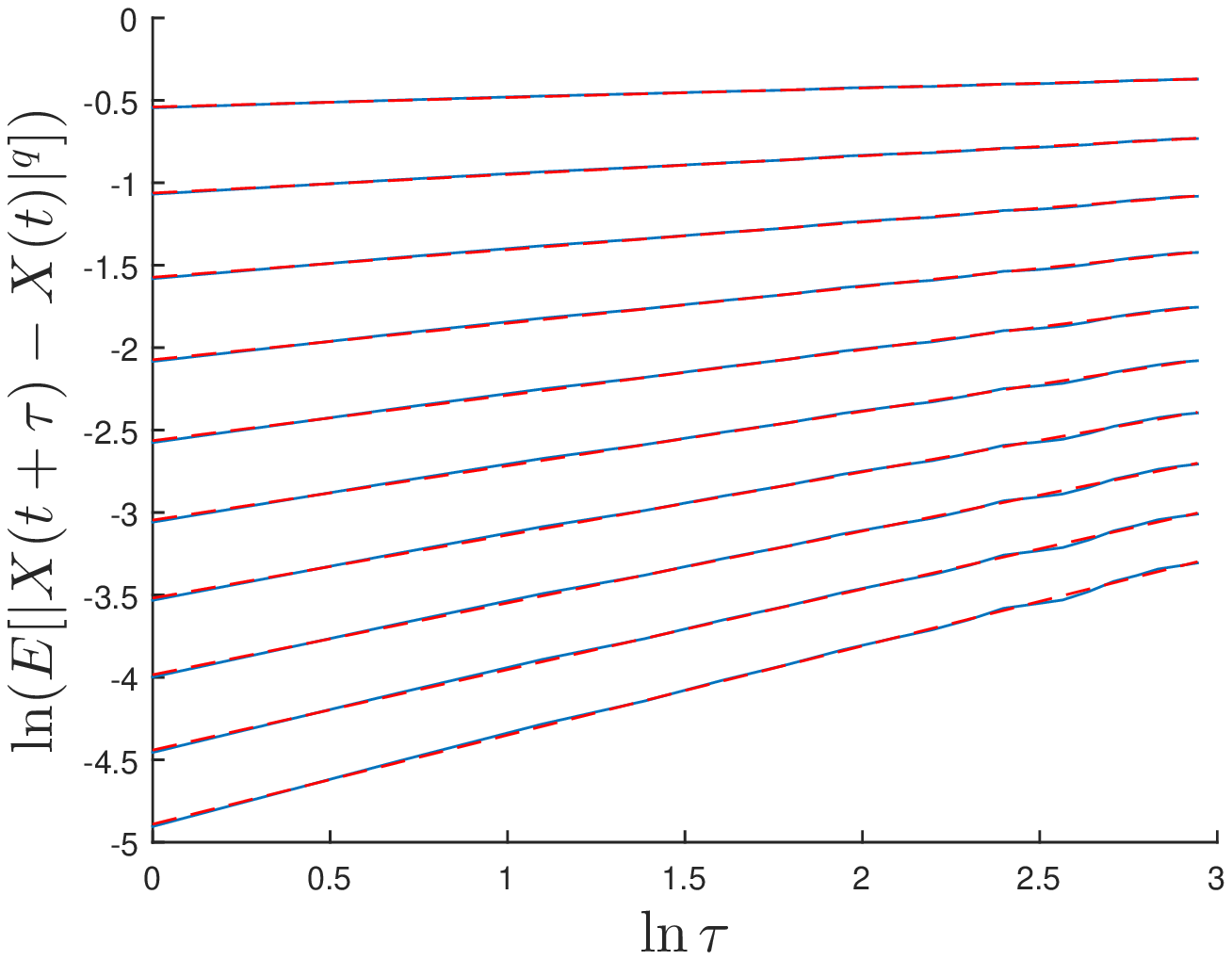}
\end{minipage}
\hfill
\begin{minipage}[b]{0.49\textwidth}
\includegraphics[width=1\textwidth]{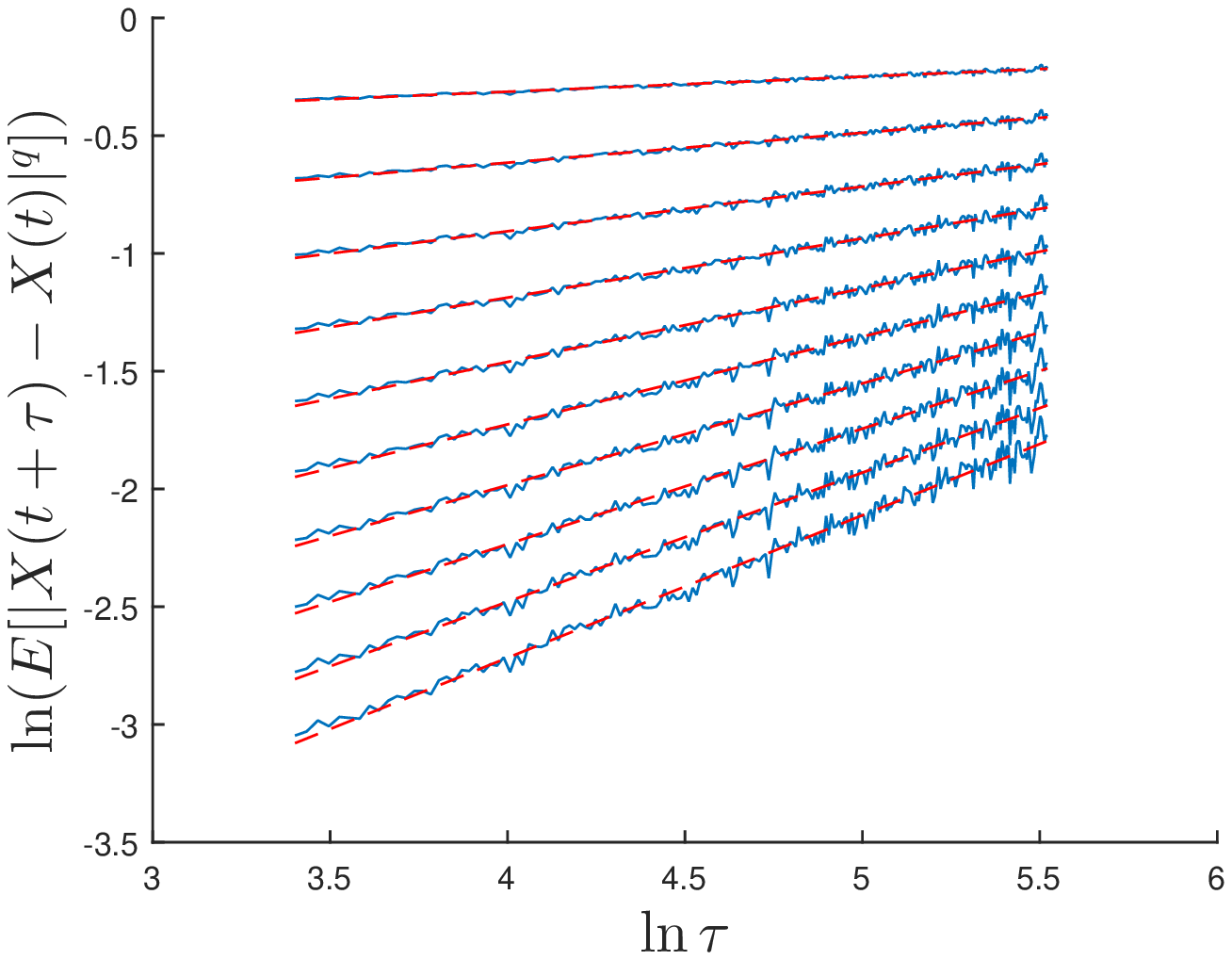}
\end{minipage}
\end{center}
\caption{Left panel: scaling of the moments of the INDU time-series with $\tau\in[1,19]$. Right panel: scaling of the moments of the INDU time-series with $\tau\in[30,250]$. The values of $q$ are taken in the interval $[0.1,1]$ every $0.1$ units, increasing from top to bottom in both panels.}
\label{indu_scaling}
\end{figure}

\begin{figure}[H]
\begin{center}
\begin{minipage}[b]{0.49\textwidth}
\includegraphics[width=1\textwidth]{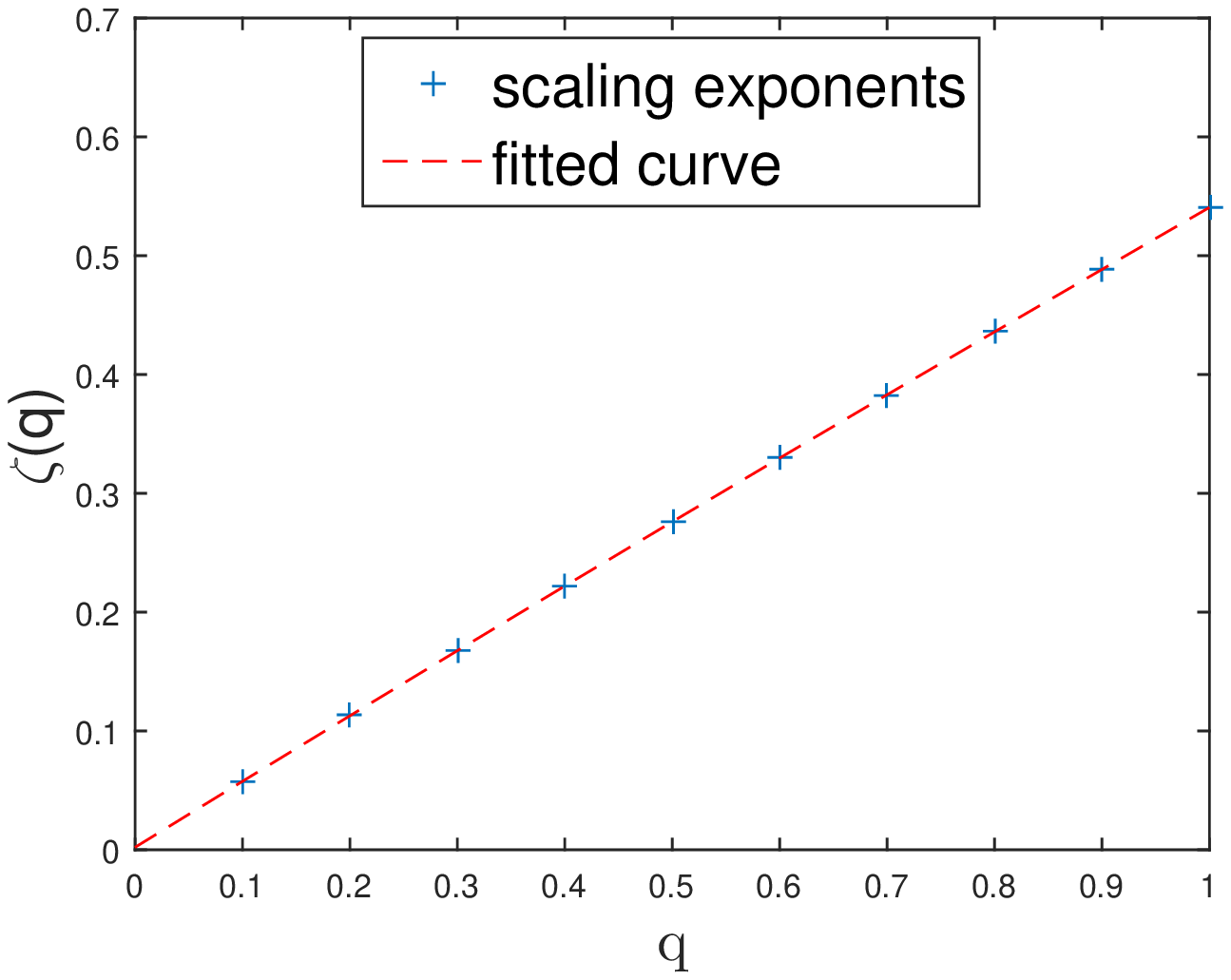}
\end{minipage}
\hfill
\begin{minipage}[b]{0.49\textwidth}
\includegraphics[width=1\textwidth]{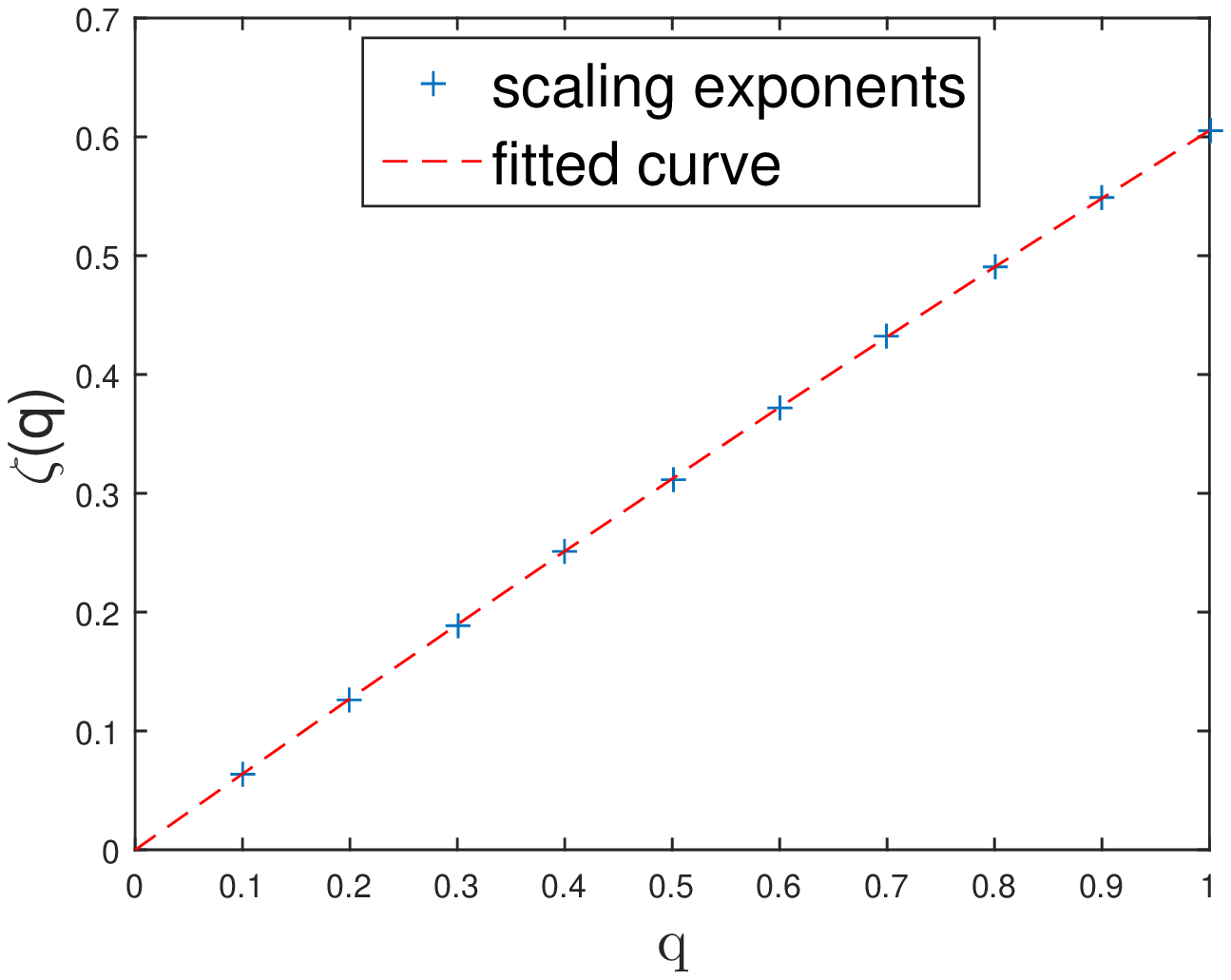}
\end{minipage}
\end{center}
\caption{Left panel: scaling exponents of the INDU time-series with $\tau\in[1,19]$. Right panel: scaling exponents of the INDU time-series with $\tau\in[30,250]$}
\label{zeta_indu_fig}
\end{figure}

The dataset we focused our attention on is the \textit{Dow Jones Industrial Average} (INDU) from 02/01/1900 to 29/12/2000 taken on a daily basis, made of $25,366$ points. We report in Fig. \ref{indu_scaling} the scaling of the moments (cfr. Eq. (\ref{moments_scaling})) respectively for $\tau\in[1,19]$ and $\tau\in[30,250]$ in blue solid lines along with their linear fit in red dashed lines.
In Fig. \ref{zeta_indu_fig} the scaling exponents $\zeta(q)$ are reported again in both regions of $\tau$, (blue crosses); as it appears evident, the parabolic shape of Eq. \ref{new_estimator} (red dashed lines) seems to fully capture the empirical behaviour.

\begin{figure}[H]
\begin{center}
\begin{minipage}[b]{0.49\textwidth}
\includegraphics[width=1\textwidth]{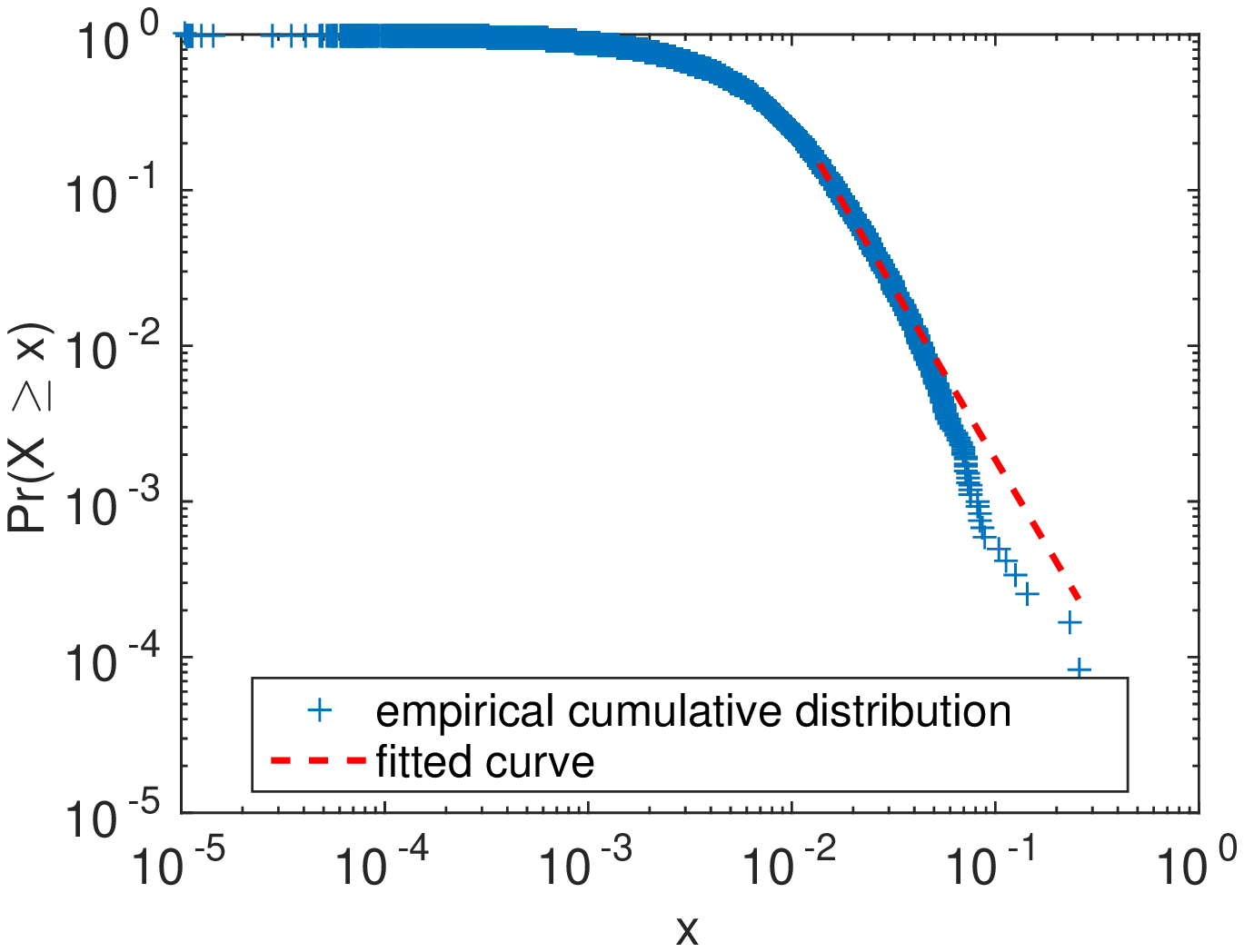}
\end{minipage}
\hfill
\begin{minipage}[b]{0.49\textwidth}
\includegraphics[width=1\textwidth]{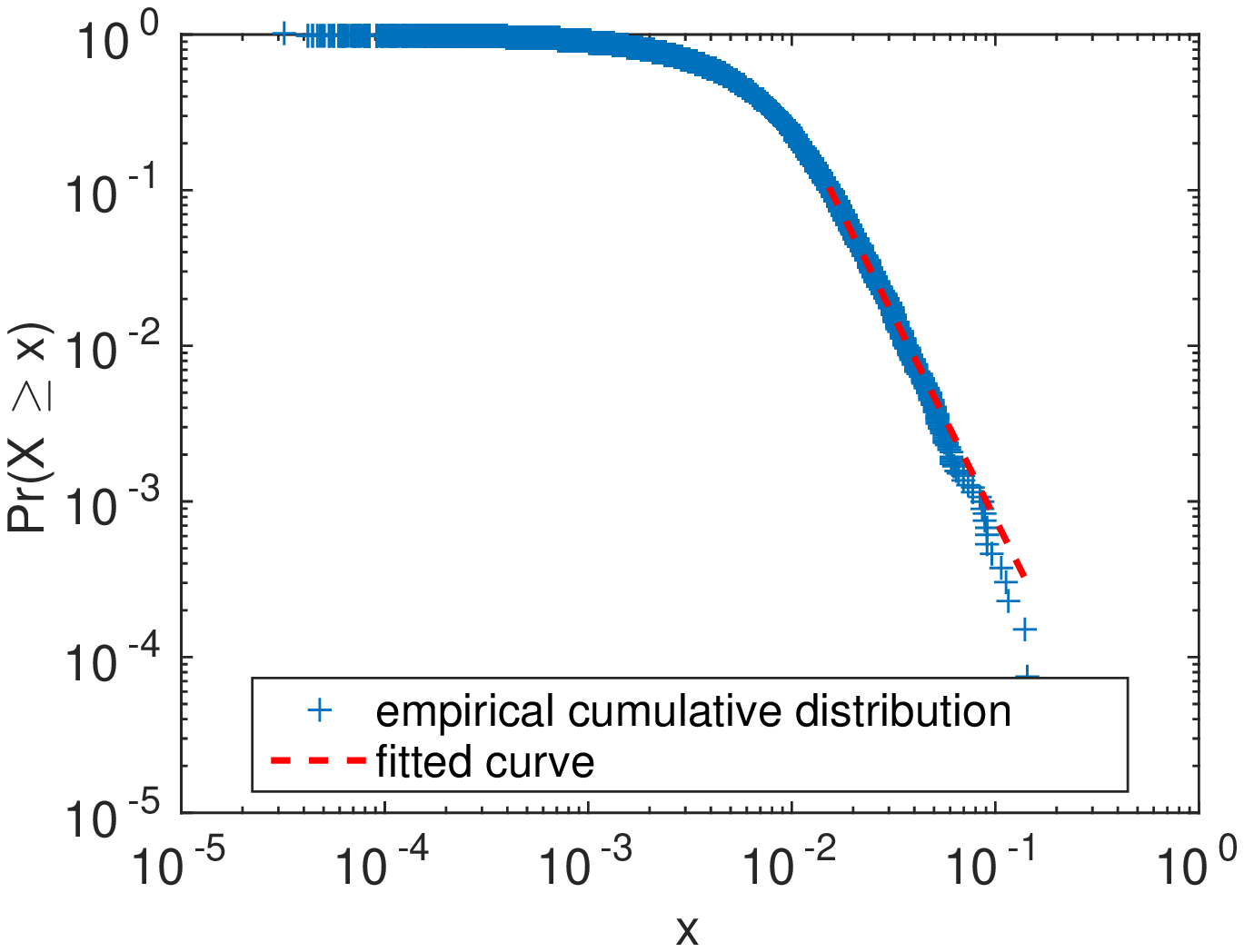}
\end{minipage}
\end{center}
\caption{Left panel: left tail of the INDU time-series. Right panel: right tail of the INDU time-series.}
\label{indu_tails_fig}
\end{figure}

This time-series exhibits power law tails and we computed the decay exponents of the tails using the method proposed in \cite{clauset1,clauset2}, based on Maximum-Likelihood Estimators and the Kolmogorov-Smirnov test. Fig. \ref{indu_tails_fig} reports the fit of the complementary cumulative distribution of the left and the right tails in loglog scale. For the left tail on the $x$-axis is reported the logarithm of minus the negative returns. The estimated values of the tails exponents are
\begin{equation}\label{indu_tails}
\begin{array}{cc}
\alpha_{left}=3.20\pm 0.05	&	\alpha_{right}=3.61\pm 0.06;
\end{array}
\end{equation}
they are different within the errors and so the time series exhibits skewness. We verified that however skewness has no effects on the measured multifractality.

\subsection{Effect of power law tails and autocorrelation in real data}
In order to uncover the source of the multiscaling behaviour of our dataset we used two procedures: the shuffling (cfr. \cite{barunik}), in order to isolate the effects of the power law tails, and the normalization (cfr. \cite{zhou}), in order to isolate the effects of the autocorrelation. We focused first on the region $\tau\in[1,19]$. A first test we made is a comparison of the scaling exponents of the INDU an a tBM, in order to check whether the empirical measured multiscaling behaviour after shuffling could be all ascribed to the presence of the power law tails or not. In order to do so, we took the INDU time-series and shuffled it $10^4$ times. On every time-series obtained we computed $\hat B$, $\hat H(0.5)$ and $\hat H(1)$, this allowed us to associate a mean and a standard deviation coming from the shuffling procedure. We then compared these values with the ones obtained computing $\hat B$, $\hat H(0.5)$ and $\hat H(1)$ on $10^4$ tBM with the same length of the INDU time-series and tails equal to the heavier empirical one, namely $\alpha_{left}$. A second test regards checking the behaviour of the INDU time-series after normalization in order to test if the change of concavity holds for empirical data. We normalized then our time-series $10^4$ times, computing the mean and the standard deviation of $\hat B$, $\hat H(0.5)$ and $\hat H(1)$.  The results are reported in Tab. \ref{indu_test_1_19} along with the value of $\hat B$, $\hat H(0.5)$ and $\hat H(1)$ computed on the plain time-series.
\begin{table}[H]
\begin{center}
\begin{tabular}{|c|c|c|c|c|}
\hline 
 					&			INDU	 	&			INDU$_{shuffled}$		&		INDU$_{normalized}$		&			tBM \\ 
\hline 
$\hat B$			&		-0.019			&			$-0.039\pm 0.003$		&		$0.0026\pm0.0005$		&		$-0.034\pm0.004$ \\ 	
\hline 
$\hat H(0.5)$		&		0.552			&			$0.572\pm0.007$			&		$0.5082\pm0.0008$		&		$0.563\pm0.007$	\\
\hline
$\hat H(1)$			&		0.541			&			$0.551\pm0.006$			&		$0.5092\pm0.0006$ 		&		$0.546\pm0.007$	\\
\hline
\end{tabular} 
\caption{Plain, shuffled and normalised INDU time-series and a t-Student with $\tau\in[1,19]$.}
\label{indu_test_1_19}
\end{center}
\end{table}
According to these simulations we confirm previous results that after shuffling the measured multiscaling behaviour of real data increases for $\tau\in[1,19]$ (see \cite{barunik}). Moreover it appears evident that this increased value is statistically undistinguishable from the one of the tBM, which is uniscaling. This result led us to infer that the multiscaling measured on shuffled empirical time-series should be ascribed only to the presence of power law tails.
\\
The normalised time-series changes its concavity after normalization (stays positive within the $1\%$ significance level), showing the same issue observed previously for the MRW.

Let us now turn our attention to the region $\tau\in[30,250]$; results are reported in Tab. \ref{indu_test_30_250}.
\begin{table}[H]
\begin{center}
\begin{tabular}{|c|c|c|c|c|}
\hline 
 					&	INDU	&	INDU$_{shuffled}$	&	INDU$_{normalized}$	&			tBM	 	\\
\hline 
$\hat B$			&	-0.038	&	$-0.01\pm 0.01$		&	$-0.0036\pm0.0007$	&		$-0.014\pm0.007$	\\ 	
\hline 
$\hat H(0.5)$		&	0.624	&	$0.53\pm0.03$		&	$0.6244\pm0.0006$	&		$0.52\pm0.02$		\\
\hline
$\hat H(1)$			&	0.605	&	$0.52\pm0.03$		&	$0.6229\pm0.0005$ 	&		$0.52\pm0.02$		\\
\hline
\end{tabular} 
\caption{Plain, shuffled and normalised INDU time-series with $\tau\in[30,250]$.}
\label{indu_test_30_250}
\end{center}
\end{table}
We observe first that the results change considerably. Secondly, within the $1\%$ significance level the shuffled time-series can be considered uniscaling, as it happens for the tBM, so there is not an increase in multifractality. Thirdly the normalized time-series  keeps its concavity, thus it is not affected anymore by the negative bias mentioned previously. This therefore demonstrates that a statistically significant multiscaling behaviour is present in financial time-series.

\section{Discussion}\label{section_discussion}
Our analyses provide clear evidence that the estimation of the scaling exponents are affected by the aggregation horizon.
We chose  two regions: 1) $\tau\in[1,19]$, which is in line with previous works and 2) $\tau\in[30,250]$. We observed that the analyses on the region $\tau\in[1,19]$ do not reproduce the theoretical expectations on time-series exhibiting power law tails or autocorrelation structures like the empirical ones. We also found an unexpected concavity of the scaling exponents $\zeta(q)$ on tBMs and nMRWs. These results are in line with previous observations on real time-series and actually enable us to give them an explanation. In particular in \cite{barunik} the authors argue that the presence of autocorrelations in real data can induce a negative bias in the estimation of the scaling exponents. According to our interpretation, the change of concavity of $\zeta(q)$ (reported in Tab. \ref{indu_test_1_19}) is exactly the effect of the negative bias. In light of this, the increased multiscaling behaviour measured in \cite{barunik} after shuffling has to be ascribed to the fact that the causal structure of a shuffled time-series is destroyed along with the negative bias itself and only the power law tails effect is left resulting in an apparent increase of multiscaling.

For what concerns the region  $\tau\in[30,250]$ we observed that the spurious multiscaling found on tBM processes and on the INDU time-series is lower with respect to the measurements performed in the $\tau\in[1,19]$ region, being even statistically absent for $n=4,5$ and for the INDU as well. Furthermore, the convexity of $\zeta(q)$ returns to a concavity, almost removing the negative bias effect. We conclude therefore that GHE measurements of multifractality in the region $\tau\in[30,250]$ are reliable and reveal that some degree of multifractality is present in real financial log-return time series and it has to be ascribed to the effect of the causal structure of the process. 

At this point a question to address is why there is a so big difference in the two regions of $\tau$. For what concerns the effect of the tails we explain this difference via the speed of convergence of the Central Limit Theorem (CLT). In particular, for processes exhibiting increments with power law tails, with tails index bigger than two, it is well-known that under aggregation they behave, in the asymptotic limit, as a BM. The speed of convergence depends on how heavy the tails are but if the aggregation is finite, whatever the tails index is, there will always be a region in the final part of the tails of the probability density which will have a power law behaviour. The effect of increasing the aggregation horizon is to push this region further in the tail. This explains why, increasing the aggregation horizon, the spurious power law tails concavity tends to disappear, reconciling with the theoretical expectations. Counter-intuitively processes with increments exhibiting tails with exponents less then two are less affected by this problem, since their convergence under aggregation is ruled by a generalized Central Limit Theorem and they keep their power law nature in the tails of the distribution so the convergence is faster. Concerning the autocorrelation negative bias, our interpretation is that it may be caused by the fact that the average of a strongly correlated variable does not necessarily converge to the expectation value. In this respect the effect might be reduced in the region $\tau\in[30,250]$ because taking bigger aggregation horizon implies averaging over less correlated variables.\\
In light of these results we argue that in order to make a reliable measure of multifractality, regions of $\tau$ with a small aggregation horizon should be taken with care. Let us however stress that the region $\tau\in[30,250]$ has not been chosen optimizing the performance of the multifractal estimator. However it proved to be sufficient to give us valuable insights and improved our estimation of the scaling parameters.

Let us make few other observations concerning the measurements. Since the measures, as proposed here (cfr. Subsec. \ref{subsec_multifractality}), depend on two parameters, $\tau_{min}$ and $\tau_{max}$, we report that in general, $\tau_{min}$ rules the precision while $\tau_{max}$ the accuracy. So a bigger value of $\tau_{min}$ would reflect in measured values nearer to expected ones. On the other hand taking bigger values of $\tau_{max}$ ends up in including more oscillating values in the analysis ,thus in a larger standard deviation. However for a process like the MRW, attention must be paid, since, if $\tau_{min}$ becomes bigger than the autocorrelation length, no multifractal behaviour holds anymore, since the increments of the process become independent. So the range of $\tau$ must be taken large enough to reduce as much as possible the power law tails effect, but not too much to exceed the time-span where the correlations are relevant. Finally, we notice that it appears evident that at small ranges of $\tau$ the power law tails concavity has a bigger impact to the measures with respect to the convexity induced by the autocorrelation.

\section{Summary and outlook}\label{section_summary}
In this paper we studied the multiscaling behaviour of  financial time-series by studying synthetic and real datasets at different aggregation horizons. We started by analysing the MRW, finding that, for small aggregation horizons, the multiscaling behaviour after shuffling, appears to increase, in agreement with previous works on empirical datasets. However for larger aggregation horizons this effect disappears. Since the shuffling procedure destroys the temporal structure of a time-series, but preserves its unconditional distribution, we focused our attention on the scaling properties of another process, the tBM which is a unifractal process. It turned out that for small aggregation horizons the presence of power law tails induces a concavity in the scaling exponents, indicating therefore a multiscaling behaviour which is however not predicted by the theory. We turned then our attention to the causal structure of a time-series. In this case we observed that, at small aggregation horizons, the presence of  autocorrelation introduces a negative bias, \textit{i.e.} a reduced concavity which ended up in a convexity of the scaling exponents, both for synthetic and real time-series. These numerical findings explain well the puzzling increase in multifractality found in previous works after shuffling: as long as both power law tails and autocorrelation are kept, the spurious multiscaling contribution of the tails is lessen by the presence of the autocorrelation, while after shuffling, only the tails effect is present. We pointed out that the aggregation of the returns is crucial. Indeed for higher aggregation horizons all these issues disappear or at least strongly lessen. For what concerns the tails we interpret this effect as a consequence of the Central Limit Theorem and its speed of convergence on time-series with power law tails but finite variance. In particular the range of tail exponents between two and five turned out to affect the most the measurements. This is due to the fact that under aggregation a residual of the power law tail is always present in the unconditional distribution and the nearer the exponent is to two, the stronger the effect. We finally note that, choosing higher values of aggregations can reduce this effect but this requires to have longer time-series. We plan in the future to study in more detail this issue trying to provide a recipe for the best choice of the region of $\tau$ which is capable to capture the multifractality of the empirical time-series.

\section*{Acknowledgements}
The authors wish to thank Bloomberg for providing the data and N. Musmeci for useful discussions. T.A. acknowledges support of the UK Economic and Social Research Council (ESRC) in funding the Systemic Risk Centre (ES/K002309/1). TDM wishes to thank the COST Action TD1210 for partially supporting this work.

\end{document}